\documentclass[11pt,a4paper]{article}

\usepackage{amsmath}
\usepackage{amsfonts}
\usepackage{amsthm}

\usepackage{graphicx}

\begin{document}

\title{Results on the Aharonov-Bohm effect without contact with the solenoid}
\author{C\'esar R. de Oliveira {\small and} Renan G. Romano\\ 
\vspace{-0.6cm}
\small
\em Departamento de Matem\'{a}tica -- UFSCar, \small \it S\~{a}o Carlos, SP, 13560-970 Brazil\\ \\}
\date{\today}

\maketitle

\begin{abstract} 
We add a confining potential to the Aharonov-Bohm model resulting in no contact of the particle with the solenoid (border); this is characterized by a unique self-adjoint extension of the initial Hamiltonian operator. It is shown that the spectrum of such extension is discrete and the first eigenvalue is found to be a nonconstant 1-periodic function of the magnetic flux circulation with a minimum  at integers and maximum at half-integer circulations. This is a rigorous verification of the effect.  
\end{abstract}



\section{Introduction and results}

Consider the Aharonov-Bohm (AB) traditional model (see the original work \cite{AharonovBohm1959}, the review \cite{Tonomura1989} and references therein for good physical accounts) of an infinitely long solenoid of radius $a> 0$, restricted to the plane, and circulation~$\kappa$ (a scaled magnetic flux inside the solenoid; see~\eqref{eqCircul}). Important theoretical issues to the Aharonov-Bohm effect are: 
\begin{itemize}
\item What is the boundary condition at the solenoid border?
\item How to confirm, through a physical quantity, the AB effect?
\item How does the AB effect depend on~$\kappa$?
\end{itemize}
Our intension is to consider a situation in which the first question above plays no role, i.e., a boundary condition is not necessary; we then confirm the AB effect (i.e., the second question above) by looking at the first eigenvalue (ground eigenvalue) $\lambda_1(\kappa)$ of the AB Hamiltonian~$H_\kappa$ (see ahead), and check a nontrivial dependence of~$\lambda_1$ on~$\kappa$. Different circulations correspond to different magnetic fields inside the solenoid, but they always vanish in the accessible region to the particle; so a nonconstant function $\lambda_1(\kappa)$ is actually a verification of the AB effect. 

Indeed, we show that $\lambda_1(\kappa)$ is a nonconstant  and even 1-periodic function, with a minimum value at integers~$\kappa$ and maximum at half-integers. Furthermore, $\lambda_1(\kappa)$ is a simple eigenvalue, except at its maxima where it is doubly degenerated, and $H_{\kappa_1}$ and $H_{\kappa_2}$ are gauge equivalent only in case $\kappa_1-\kappa_2$ is an integer number (the period of $\lambda_1(\kappa))$. In this work we report such results  obtained on a rigorous mathematical basis and, importantly, describe the main ideas behind them;  proofs may be found in~\cite{dOR}.

To be more precise, we begin with some notations: the accessible region is $\Omega_{a}:=\left\{z\in\mathbb{R}^{2}:\left|z\right|>a\right\},$ $z=\left(x,y\right)$ so that $r=\left|z\right|=\sqrt{x^{2}+y^{2}}$. We also denote the (usual) AB magnetic potential in~$\Omega_a$, for a real parameter~$\kappa$,
\begin{equation*}\label{eq:Akappa}
	\mathbf{A}_{\kappa}\left(z\right):=\frac{\kappa}{\left|z\right|^{2}}\left(-y,x\right).
\end{equation*}
Note that $\mathbf A_\kappa$ is smooth and bounded.

We consider an additional confining scalar (radial) potential $V(r)$ in $\Omega_{a}$, positive, continuous and satisfying
\begin{equation}\label{eq:defV1231}
	V\left(r\right)\geq\frac{1}{\left(r-a\right)^{2}}
\end{equation}
	for $r$ close to the solenoid border $r=a$, and finally
\begin{equation}\label{VdivInf}
	\lim_{r\rightarrow\infty}V\left(r\right)=\infty.
\end{equation}

Our initial Aharonov-Bohm operator, in suitable units,  is then 
\begin{equation*}\label{eq:ABdefmeu}
	\dot H_{\kappa}\varphi:=\left(i\nabla+\mathbf{A}_{\kappa}\right)^{2}\varphi+V\varphi,
\end{equation*}
for the usual smooth functions of compact support $\varphi\in \mathrm{C}_{0}^{\infty}\left(\Omega_{a}\right)$. Any self-adjoint extension of $H_{\kappa}$ is a fair candidate to model the AB effect. However, by using methods discussed in \cite{NenciuNenciu2009,DeVerdiereTruc2010}, we are able to show that condition \eqref{eq:defV1231} implies that $\dot H_\kappa$ has just one self-adjoint extension $H_\kappa$, which then is the operator describing the modeling in this situation. The lack of  boundary conditions at the solenoid border is interpreted as no contact of the particle in~$\Omega_a$ with the solenoid.

By using the technique described in \cite{lewis}, it is obtained from \eqref{VdivInf} that the spectrum of $H_\kappa$ is always discrete. We then try to use the ground state eigenvalue $\lambda_1(\kappa)$ to probe the presence of the AB effect, that is, whether $\lambda_1(\kappa)$ is a nonconstant function of~$\kappa$ despite the fact that $\nabla\times \mathbf{A}_{\kappa}=0$  (i.e., zero magnetic field in~$\Omega_a$). This idea has already been considered in other situations \cite{helffer,HelfferHoffman1999}.

Given a smooth magnetic potential $\mathbf{A}$ in~$\Omega_{a}$ with $\nabla\times\mathbf{A}=0$, define its \textsl{circulation} as
\begin{equation}\label{eqCircul}
	\kappa\left(\mathbf{A}\right):=\frac{1}{2\pi}\int_{{\mathcal C}}\mathbf{A}\cdot\mathrm{ds},
\end{equation}
for any circumference ${\mathcal C}$ in~$\Omega_{a}$  centered at the origin. Stokes' Theorem guarantees that the above integral has the same value for all such circumferences. The circulation of $\mathbf{A}_{\kappa}$ is exactly~$\kappa$, so justifying the abuse of notation.

 Two general smooth magnetic potentials $\mathcal{A}_{1}$ and $\mathcal{A}_{2}$ in~$\Omega_{a}$ are said to be  \textit{gauge classically related} if there exists a real smooth  multivalued function $\phi$ in~$\Omega_{a}$ so that $\nabla \phi$  is   well defined and
\begin{equation*}\label{eq:potiq}
	\mathcal{A}_{2}=\mathcal{A}_{1}+\nabla\phi\quad\textrm{ in }\Omega_{a}.	
\end{equation*}
In  particular $\nabla \times \mathcal{A}_{1}= \nabla \times \mathcal{A}_{2}$ (same magnetic field). If, furthermore, $e^{i\phi}$ is also single valued, we say that $\mathcal{A}_{1}$ and $\mathcal{A}_{2}$  \textit{gauge quantum related} (GQR). It is possible to show that  $\mathcal{A}_{1}$ and $\mathcal{A}_{2}$ are GQR if, and only if,  they are gauge classically related and for all circumferences ${\mathcal C}$ in~$\Omega_a$, one has
\begin{equation*}\label{eq:gaugeeq}
	\left(\dfrac{1}{2\pi}\int_{{\mathcal C}}\left(\mathcal{A}_{2}\left(z\right)-\mathcal{A}_{1}\left(z\right)\right)\cdot\mathrm{d}s\right)\in\mathbb{Z},
\end{equation*}
that is, $\kappa(\mathcal A_2)-\kappa(\mathcal A_1)$ is an integer number. It then follows that $\mathbf A_\kappa$ and $\mathbf A_{\kappa'}$ are GQR if, and only if, $\kappa-\kappa'$ is an integer; in particular, in the interval $(-1/2,1/2]$ there is no pair of circulations $\kappa, \kappa'$ so that $\mathbf A_\kappa$ and $\mathbf A_{\kappa'}$ are gauge related. It is also possible to show that $\mathbf A_\kappa$ and $\mathbf A_{\kappa'}$ are GQR is equivalent to the fact that the corresponding Hamiltonian operators $H_\kappa$ and $H_{\kappa'}$ are \textit{gauge equivalent}, that is, there exists a real smooth multivalued function $\phi$ in~$\Omega_a$ so that  $e^{i\phi}:\Omega_{a}\rightarrow\mathbb{C}$  is  single valued (so defining a unitary operator, the so-called gauge transformation) with $\mathrm{dom}\,{H}_{\kappa'}=e^{i\phi}\mathrm{dom}\,{H}_{\kappa}$ and 
\[
{H}_{\kappa'}e^{i\phi}=e^{i\phi}{H}_{\kappa}.
\] Summing up, for any integer number $m$, $H_\kappa$ is gauge (unitarily)  equivalent to $H_{\kappa+m}$, and such operators are not gauge equivalent for any $\kappa,\kappa'\in(-1/2,1/2]$, $\kappa\ne\kappa'$. With respect to spectrum, in particular the first eigenvalue $\lambda_1(\kappa)$, it is then a periodic function of period~1, and we may restrict ourselves to~$\kappa$ in the interval $(-1/2,1/2]$.

On the other hand, it is clear that $T\dot H_\kappa=\dot H_{-\kappa}T$, where $T$ is the complex conjugation  $T\psi:=\overline\psi$, an antiunitary operator; by extension one gets $T H_\kappa= H_{-\kappa}T$. By using Weyl sequences \cite{DeOliveira2000}, one sees that $H_\kappa$ and $H_{-\kappa}$ have the same spectrum (but they are not gauge equivalent; it is an open question whether they are unitarily equivalent or not). By combining with the above conclusion, it follows that $\lambda_1(\kappa)$ is an even 1-periodic  function. We may then restrict ourselves to~$\kappa$ in the interval $[0,1/2]$.

It is still missing a crucial ingredient, that is, to check that despite the zero magnetic field in~$\Omega_a$, for all~$\kappa$, the first eigenvalue is a nonconstant function, so characterizing the AB effect in this situation. 

\begin{figure}[htbp]
\begin{center}
\includegraphics[width=11.52cm, height=7.68cm]{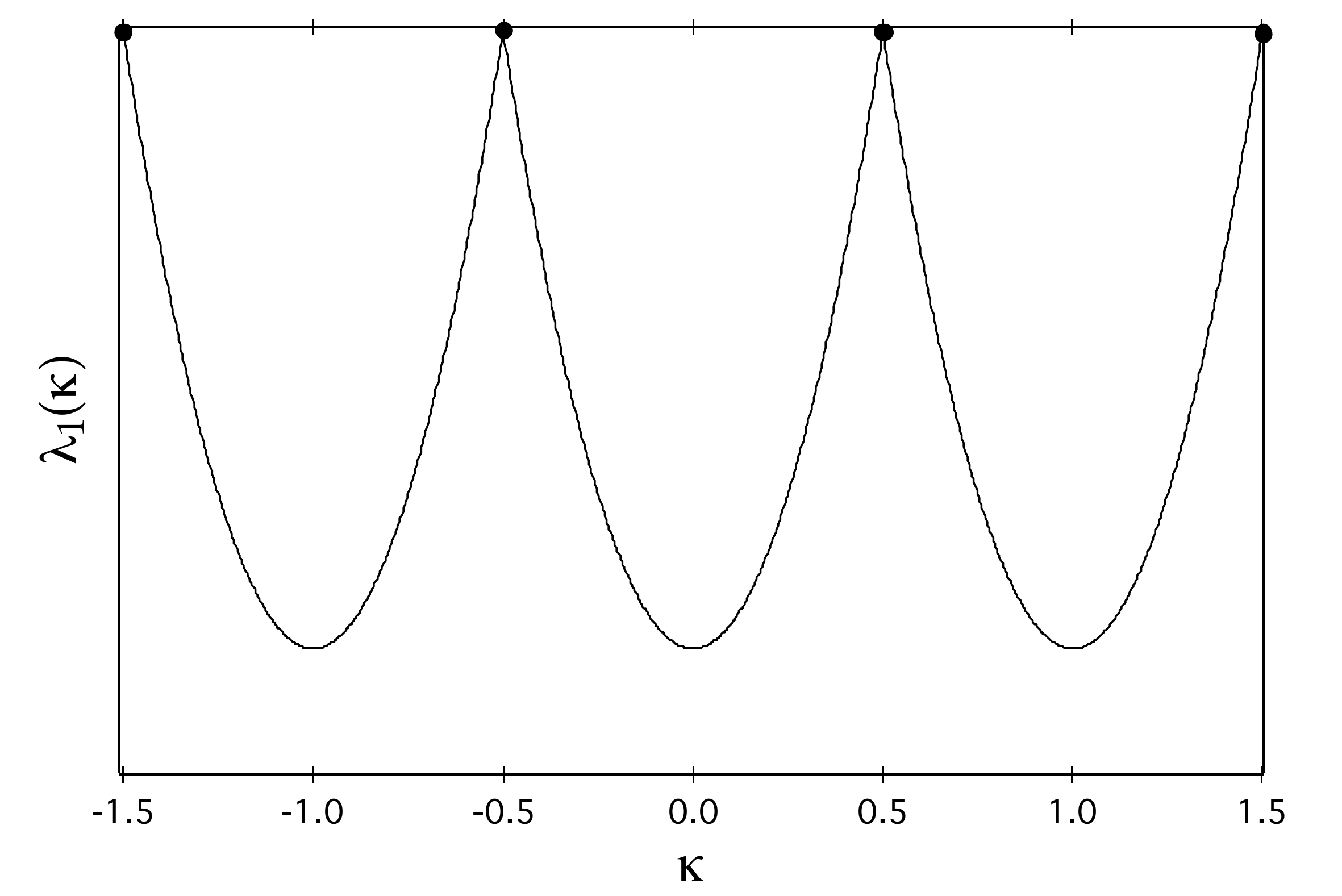}
\caption{Behavior of the first eigenvalue $\lambda_1(\kappa)$ as function of the circulation~$\kappa$. Note the periodicity and the maximum values (doubly degenerated) indicated by full circles.}
\label{labelFig}
\end{center}
\end{figure}

It is found that $\lambda_1(\kappa)$ is nondegenerate in $[0,1/2)$, and by applying the theory of analytic perturbations and sectorial forms~\cite{12kato}, one checks that  $\left(H_{\kappa}\right)_{\kappa\in\left(0,1/2\right)}$ is a self-adjoint analytic family of Type~$(B)$ with compact resolvent (see Remark~4.22, Section~4, Chapter~7 in~\cite{12kato}), so that $\lambda_1(\kappa)$ as well as the corresponding eigenfunctions $\psi_\kappa$ are analytic ($H_\kappa\psi_\kappa=\lambda_1(\kappa)\psi_\kappa$). A direct computation with the actions of the operators gives
\begin{equation*}\label{eqDerivAutov}
\dfrac{\mathrm{d}\lambda_1(\kappa)}{\mathrm{d}\kappa}\left\|\psi_\kappa\right\|^{2}=\left\langle 2i\mathbf{A}_{1}\cdot\nabla \psi_\kappa,\psi_\kappa\right\rangle+2\kappa\left\|\dfrac{\psi_\kappa}{\left|z\right|}\right\|^{2},
\end{equation*} and since in this range of $\kappa$  it is possible to choose the fundamental state $\psi_\kappa$ real-valued and normalized, one has
\[
\dfrac{\mathrm{d}\lambda_1(\kappa)}{\mathrm{d}\kappa}=2\kappa\left\|\frac{\psi_\kappa}{\left|z\right|}\right\|^{2}>0,\quad \kappa\in (0,1/2).
\]
It  follows that the first eigenvalue is an increasing function of~$\kappa$ in $[0,1/2]$,  and so with minimum at $\kappa=0$ (no magnetic field inside the solenoid), as well as at all integer values by periodicity, and maxima at half-integers; see Figure~\ref{labelFig}. This is a rigorous (theoretical) verification of the existence of the AB effect in this context with no contact with the solenoid.

Final remarks. (a)~The period of $\lambda_1(\kappa)$ is compatible with experimental scattering results \cite{Tonomura1989} (without the confining~$V$). (b)~By using the decomposition of~$H_\kappa$ in polar coordinates, it is also found that $\lambda_1(\kappa)$ has multiplicity 2  at such maximum values, and in this case  there is an eigenfunction with a nodal set that is a simple line, starting at the solenoid border and going to infinity. (c)~$\dot H_\kappa$  has a unique self-adjoint extension for  solenoids of more general shapes than disks. (d)~The same conclusions hold true for radius $a=0$ solenoids, but under more specific choices of additional potentials~$V$. (e)~It is worth mentioning that an important technical point in the proofs is the quadratic form~$\mathcal L_\kappa$ associated with $H_\kappa$, that is,
\begin{equation*}\label{eq:sesqlinform}
	\mathcal L_{\kappa}\left(\psi\right):=\int_{\Omega_{a}}\left(|(i\nabla+\mathbf{A}_{\kappa})\psi|^2+V|\psi|^2\right)\mathrm{d}x\mathrm{d}y,
\end{equation*}with domain $\mathcal H_0^1(\Omega_a)\cap \mathrm{dom} V^{1/2}$ ($\mathcal H_0^1$ is the usual Sobolev space). (f)~The case $a=0$ with no additional potential~$V$ and Dirichlet condition is known~\cite{pankrRich} to have absolutely continuous spectrum $[0,\infty)$.

\subsection*{Acknowledgments}
CRdO thanks CNPq for partial support (Universal Project 41004/2014-8). RGR was supported by FAPESP (under contract 2012/21480-8).


\begin{thebibliography}{99}

\bibitem{AharonovBohm1959}Y. Aharonov and D. Bohm, {Phys. Rev.} \textbf{115} 485--491 (1959).

\bibitem{Tonomura1989}M. Peshkin and A. Tonomura,  The Aharonov-Bohm Effect. LNP340. Berlin: \textsl{Springer-Verlag} (1989).

\bibitem{dOR} C. R. de Oliveira and R. G. Romano, J. Math. Phys. \textbf{58} 102102 (2017).



\bibitem{NenciuNenciu2009}G. Nenciu and I. Nenciu, {Ann. H. Poincar\'{e}} \textbf{10} 377--394 (2009).

\bibitem{DeVerdiereTruc2010}Y. Colin  de Verdi\`{e}re and F. Truc,  Ann. Inst. Fourier  \textbf{60} 2333--2356 (2010).

\bibitem{lewis}R. T. Lewis, {Trans. Amer. Math. Soc.} \textbf{271} 653--666 (1982).

\bibitem{helffer}B. Helffer, {Commum. Math. Phys.} \textbf{119} 315--329 (1988).
	
\bibitem{HelfferHoffman1999}B. Helffer,  M. Hoffmann-Ostenhof, T. Hoffmann-Ostenhof and M. P. Owen, {Commun. Math. Phys.} \textbf{202} 629--649 (1999).

\bibitem{DeOliveira2000}C. R. de Oliveira,  \textit{Intermediate Spectral Theory and Quantum Dynamics} (Birkh\"auser, Basel, 2009).

\bibitem{12kato}T. Kato,  \textit{Perturbation Theory for Linear Operators}, Second Edition (Spring-Verlag,  Berlin, 1979).

\bibitem{pankrRich} K. Pankrashkin and S. Richard,  Rev. Math. Phys. \textbf{23} 53--81 (2011).

\end{thebibliography}
\end{document}